\documentclass[10pt,conference]{IEEEtran}
\IEEEoverridecommandlockouts
\usepackage[letterpaper, left=0.653in, right=0.673in, top=0.793in, bottom=1in]{geometry}
\usepackage{cite}
\usepackage{amsmath,amssymb,amsfonts} 
\usepackage{graphicx}
\usepackage{textcomp}
\usepackage{xcolor}
\usepackage{algorithm}
\usepackage{algpseudocode}
\usepackage{booktabs} 
\usepackage{multirow}

\def\BibTeX{{\rm B\kern-.05em{\sc i\kern-.025em b}\kern-.08em
    T\kern-.1667em\lower.7ex\hbox{E}\kern-.125emX}}
\begin{document}

\title{Safety-Tiered ISAC Scheduling with History-Aware Value of Information for V2X Networks
}
\author{
Mubashar~Sarfraz,
Nima~Imani,
Spyros~Lavdas,
and~Constantinos~B.~Papadias%
\thanks{This work was supported by the EU MSCA Doctoral Network ISLANDS under grant agreement No.~101120544.}%
\thanks{Mubashar Sarfraz, Nima Imani, Spyros Lavdas, and Constantinos B. Papadias are with the SWiFT Lab, The American College of Greece, Research Center, Athens 15342, Greece.}%
\thanks{Mubashar Sarfraz and Nima Imani are with the Department of Electrical and Computer Engineering, University of Cyprus, Nicosia 1678, Cyprus.}%
\thanks{Corresponding author: Mubashar Sarfraz (e-mail: msarfraz@acg.edu).}%
\thanks{\copyright~2026 IEEE. Personal use of this material is permitted. Permission from IEEE must be obtained for all other uses, in any current or future media, including reprinting/republishing this material for advertising or promotional purposes, creating new collective works, for resale or redistribution to servers or lists, or reuse of any copyrighted component of this work in other works. Accepted for publication in the 2026 IEEE 1st Annual Integrated Sensing and Communication Conference (ISAC), Lisbon, Portugal, 16--18 Nov. 2026.}%
}

\maketitle
\thispagestyle{empty}
\begin{abstract}
Integrated sensing and communication (ISAC) assisted vehicle-to-everything (V2X) networks prioritize safety-critical radar sensing over communication. However, existing scheduling methods overlook the varying criticality of vehicles, long-term fairness in service quality, and information staleness. Considering this, we propose a low-complexity sensing-prioritized allocation for radar and communication (SPARC-V) to maximize the value of information (VoI) in V2X  networks. SPARC-V addresses these challenges with a two-stage hierarchical resource scheduling framework. For sensing, vehicles are first categorized into safety tiers based on a risk metric. Safety-critical vehicles receive orthogonal resources, while others can reuse resources under specific constraints. Sensing resources are allocated based on a sensing priority metric that balances safety significance and long-term fairness. In the next stage, communication resources are allocated to vehicles with the most outdated information. Performance is measured using a unified VoI metric that considers sensing quality, safety relevance, and information freshness. Monte Carlo simulations show that SPARC-V achieves the highest network VoI of 26.5 among all baselines, with its advantage over baselines widening as vehicle density increases.
\end{abstract}

\section{Introduction}
\label{sec:intro}
Integrated sensing and communication (ISAC) enables radar-based vehicle detection and cooperative awareness message (CAM) transmission in vehicle-to-everything (V2X) networks using unified spectrum and hardware \cite{zhang2026integrated}. CAMs are  periodically transmitted to share real-time situational information with roadside units (RSUs) and other vehicles \cite{11219062}.  Stale or missing CAM and sensing data from a rapidly approaching vehicle can directly increase the risk of collision. 
\par
The age of information (AoI) addresses staleness in status-update networks \cite{11396674}, and AoI-based scheduling is well established \cite{AoI}. The value of information (VoI) extends it to the decision-theoretic relevance of information, prioritizing safety-critical data for collision avoidance \cite{wang2024TVT,rolich2026latency}. Both metrics are instantaneous, however: a vehicle repeatedly denied adequate sensing accumulates no priority and cannot recover, and a vehicle with a fresh CAM but inadequate sensing still yields unreliable collision-risk estimates. ISAC-V2X resource allocation typically balances communication rate against sensing accuracy \cite{liu2020radar}, or maximizes energy efficiency under detection-probability constraints \cite{11594737}, without addressing long-term fairness. To the best of our knowledge, no prior work integrates (i) mobility-aware safety tiering of vehicle pairs, (ii) long-term sensing quality history (SQH), and (iii) communication freshness.
\par
Closest to this work, Abedi et al.~\cite{abedi2026precise} propose a safety-aware, VoI-driven ISAC framework for human-driven vehicle positioning, solved via multi-agent deep reinforcement learning. That approach requires offline training and policy convergence, and its VoI, like those above, carries no record of the sensing quality delivered over time; SPARC-V instead allocates in a single greedy pass within the 25~ms scheduling period.
\par
This paper presents SPARC-V, a low-complexity two-stage scheduler for ISAC-assisted V2X. We introduce a velocity-aware risk metric that tiers vehicle pairs into critical and cautionary classes, an SQH metric that records delivered detection energy over time, and a sensing priority (SP) metric that combines the two so that historically underserved vehicles regain priority. Residual resources are then allocated for uplink CAM transmission in descending order of AoI. Performance is evaluated with a unified VoI metric capturing sensing quality, safety relevance, and freshness.
\section{System Model}
In this paper, we consider a single RSU that acts as a centralized access point and scheduler in an ISAC-assisted V2X network. The RSU manages 
resource distribution among $N$ ISAC-equipped vehicle users ($\mathcal{V}$), where each $V_i$, 
$i\in\{1,2,\ldots,N\}$, is distributed along a straight road of length $L$ with two parallel lanes separated by a fixed lateral offset. The position of each vehicle is denoted $\mathbf{q}_i \in \mathbb{R}^{1\times 2}$, and the RSU is located at the road midpoint $q_{\text{RSU}}$. The RSU 
forms valid ISAC pairs based on proximity within each lane: vehicles $V_i$ and $V_{i+1}$
constitute a valid sensing pair if their inter-vehicle distance $d_i^r=\|\mathbf{q}_{i+1}-\mathbf{q}_i\|_2\leq d_{max}^r$, where $d_{max}^r$ is the maximum detection range.
Pairs exceeding this threshold are ineligible for sensing and no radar resources are allocated, 
defining the sensing-eligible set $\mathcal{V}_{\mathrm{sense}}=\{i\in\mathcal{V}:d_i^r\leq d_{max}^r\}$. To limit mutual V2V interference under non-orthogonal resource sharing, a minimum reuse distance $d_{sep}^{min}$ is enforced, ensuring that any two pairs sharing the same time-frequency resources are spatially separated.
\par
The system follows the 3GPP NR numerology with a subcarrier spacing (SCS) of 60 kHz. Each scheduling period lasts for $T_{\mathrm{total}} = 25$~ms, utilizing 25 resource blocks, with each resource block spanning 1~ms and containing $N_{total} = 56$ OFDM symbols. Each of these resource blocks contains 4 time slots, which are further subdivided into $N_{mini} = 8$ mini-slots per block, each containing $N_{sym} = 7$ symbols and lasting $T_{slot} = 0.125$ ms. 
Each vehicle $V_i$ operates in time-switching mode within each scheduling round. It performs radar sensing in its allocated sensing mini-slots by transmitting toward $V_{i+1}$ and processing the backscattered echo, then transmits its CAM to the RSU in its communication mini-slots. The line-of-sight (LoS) path loss from $V_i$ to the RSU is $\mathcal{L}_{\mathrm{LoS}}(d^c)=21.5\log_{10}(d^c)+44.2+20\log_{10}(f_c)$~dB, with $d^c$ in m and $f_c$ in GHz. Non-line-of-sight (NLoS) path loss is $\mathcal{L}_{\mathrm{NLoS}}(d^c) = \mathcal{L}_{\mathrm{LoS}}(d^c)
+ \mathcal{L}_{\mathrm{NLoS}}^o+ \chi$, where $\mathcal{L}_{\mathrm{NLoS}}^o=20$~dB is the fixed additional attenuation,  $\chi \sim N(0,\sigma_\chi^2)$  is a zero-mean Gaussian shadowing term with standard deviation $\sigma_\chi$ and each vehicle is independently classified as NLoS with probability $P_{NLoS}$ \cite{3GPPNLoS}.

\par
\begin{figure}[t]
\centering
\includegraphics[width=\columnwidth]{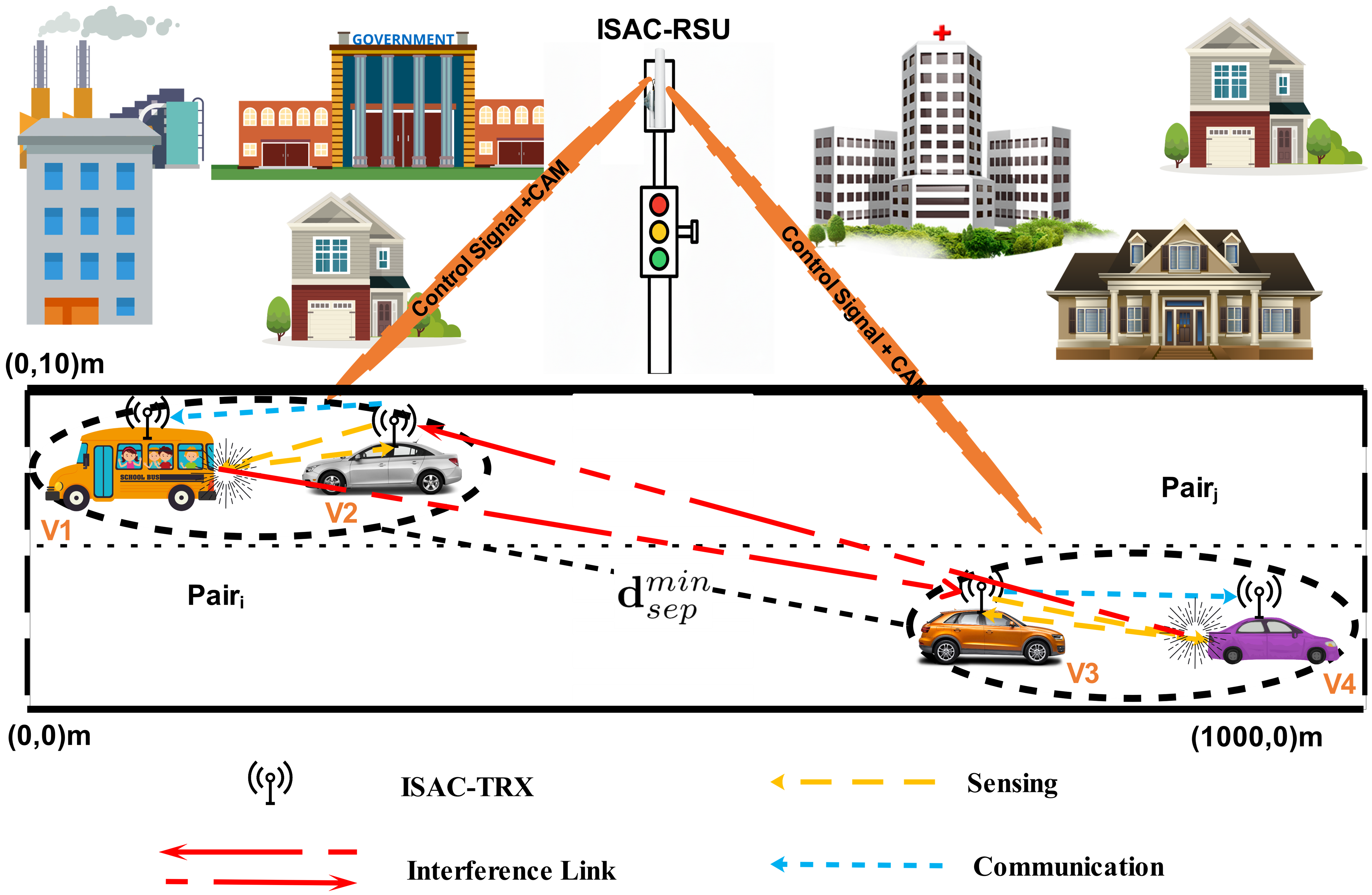}
\caption{ISAC-V2X network illustrating sensing, communication, and interference links for vehicle pairs with a minimum reuse distance $d_{sep}^{min}$.}
\label{system_model}
\vspace{-5mm}
\end{figure}
The OFDM waveform occupies bandwidth $B$ with SCS $\Delta f$, giving $N_{used}=\lfloor B/\Delta f\rfloor$ subcarriers and occupied bandwidth $B_{used}=N_{used}\,\Delta f$. Distinct noise figures $F_{VU}$ and $F_{RSU}$ apply to the sensing and communication links, giving noise powers $\sigma^2_{VU}=k_B T_0\,10^{F_{VU}/10}B_{used}$ and $\sigma^2_{RSU}$ likewise, with $T_0=300$~K.
\par
Given the false alarm probability $P_{fa}$, the detection probability $P_d$ for the Rayleigh fluctuating target \cite{de2008design} can be calculated as 
\begin{equation}
P_d
=
\exp\!\left(
\frac{\ln P_{fa}}{1+\gamma^r_i}
\right),
\label{eq:Pd}
\end{equation}
\par
Setting $P_d = P_{d,Req}$ in \eqref{eq:Pd} and solving for $\gamma_i^r$ yields $\gamma^r_{\min}
=\frac{\ln\!\left(P_{\mathrm{fa}}\right)}
{\ln\!\left(P_d\right)}
- 1.$
This gives the total detection SINR required over the sensing burst. As in \cite{11594737}, coherent integration over the granted symbols accumulates the per-symbol SINR of \eqref{eq:MI}, so a vehicle allocated $N_i^{\mathrm{sym}}(t)$ symbols meets the detection requirement when $\gamma_i^r(t) N_i^{\mathrm{sym}}(t) \ge \gamma_{\min}^r$. The corresponding per-symbol target is $\gamma_{\min,1}^r = \gamma_{\min}^r / N_i^{\mathrm{sym}}$.
\par
The sensing quality is determined by sensing SINR which is defined as 
\begin{equation}
\gamma^r_i = \frac{
P_{i}^r \, \lvert h_{i}^r \rvert^{2}
}{
\sum_{\substack{j \\ j \neq i}}
z_{i,j} \ P^r_{j} \, \lvert h^{v2v}_{j \rightarrow i } \rvert^{2}
+\sigma^2_{VU}}
\label{eq:MI}
\end{equation}
where $P^r$ is the radar power, $h^{r}_i$ = $\frac{\lambda G \sqrt{\sigma_r}\, }
{(4\pi)^{3/2} (d^r_i)^{2}}$ is the two way radar channel, $h^{v2v}_{j \rightarrow i }  = \lambda G^{0.5} 
/ (4\pi d_{ji}) $ is the one-way vehicle-to-vehicle (V2V) channel from interfering vehicle $V_j$ at distance $d_{ji}$, $\lambda$ is the wavelength, $G$ is the antenna gain, $\sigma_r$ is the vehicle's cross section and $z_{i,j}$ is the reuse indicator indicating whether the resources for the $i^{th}$ and $j^{th}$ users are orthogonal ($z_{i,j}=0$) or non-orthogonal ($z_{i,j}=1$).
\par
The communication channel from vehicle $V_i$ to the RSU is $h_i^c = 10^{(G_c-\mathcal{L}(d_i^c))/10}$, where $G_c$ is the aggregate communication antenna gain in dB. The target SINR threshold $\gamma^c_{target}(M_i)$ for modulation and coding scheme (MCS) index $M_i$ is calculated from the CAM payload, the 3GPP NR resource element structure, and a truncated Shannon bound with attenuation factor $\kappa=0.4$ \cite{3GPPMCS,3GPPalpha}.
\par
The risk metric $\rho_i$, used as a risk score rather than a time-to-collision estimate, captures the urgency of sensing in a mobility-aware manner by jointly considering the current inter-vehicle distance $d_i^r$ and the relative speed between vehicle $V_i$ and its paired target is given as
\begin{equation}
\rho_i(t)=\frac{d_i^r}{v_{rel,i}+\varepsilon_v}
\label{eq:risk}
\end{equation}
where $v_{rel,i} = |v_i - v_{i+1}|$ is the relative speed between the pair, and 
$\varepsilon_v > 0$ is a small stability constant preventing division by zero when vehicles travel at 
equal speeds. A low value of $\rho_i$ indicates a high collision risk and it necessitates the allocation of orthogonal sensing resources. A high $\rho_i$ indicates lower urgency and permits resource reuse.
\par
Vehicles are classified into two safety tiers based on the risk metric \( \rho_i \). A vehicle pair is considered critical if its risk metric falls below the safety threshold \( \rho_{th} \). This situation requires the use of orthogonal, interference-free sensing resources. In contrast, if the risk metric is above the threshold, the vehicle pair is classified as cautionary and may engage in non-orthogonal resource reuse. This classification is represented by a binary indicator $\delta_i, \ \text{equal to 1} \ \text{if } \rho_i > \rho_{th}$ (cautionary) and
$0 \ \text{otherwise} \ \text{(critical)}$. The collision risk is mapped to a bounded safety relevance function $g(\rho_i(t)) = \exp\left( -\rho_i(t)/ \rho_{th} \right)$.
\par
The AoI for vehicle $V_i$ counts the number of scheduling rounds since the RSU last successfully decoded its CAM. It is initialized as $AoI_i(0) = 1$, which is  consistent with the discrete-time AoI formulations in \cite{AoI} and evolves as follows
\begin{equation}
\mathrm{AoI}_i(t+1) = \begin{cases} 1, & \text{if CAM decoded} , \\ \mathrm{AoI}_i(t) + 1, & 
\text{otherwise.} \end{cases}
\label{eq:AoI2}
\end{equation}
\par
The corresponding physical age is $\Delta_ i(t)=\mathrm{AoI}_i(t) T_{total}$, so the discrete index in \eqref{eq:AoI2} is a quantized representation of continuous age with resolution $T_{total}$ = 25 ms. The initialization $\mathrm{AoI}_i(0) = 1$ therefore corresponds to a minimum attainable age of 25 ms, set by the scheduling period rather than by the policy.
\par
A vehicle \( V_i \) is successfully decoded if it receives communication resources and its SINR meets the MCS target, i.e., 
$\text{decoded}_i(t) = \mathbf{1}[s_i^c(t) > 0]  \mathbf{1}[\gamma_i^c(t) \geq \gamma_{target}^c(M_i)].$ At each round, the RSU ranks the vehicles in 
descending order of $AoI_i$ and assigns communication resources to the vehicles that are the most stale 
first. The SQH metric ensures long-term fairness in sensing resource allocation. This metric tracks the quality of sensing received by each vehicle, which is calculated as  
\begin{equation}
\mathrm{SQH}_i(t+1)=\frac{t\,\mathrm{SQH}_i(t)+\min\!\left(1,\frac{\gamma_i^r(t)N_i^{\mathrm{sym}}(t)}{\gamma_{\min}^r}\right)}{t+1}
\label{eq:SQH}
\end{equation}
\par
The per-round credit is the integrated detection energy delivered relative to the requirement. Deferred vehicles receive $N_i^{\mathrm{sym}} = 0$ and contribute zero; served vehicles contribute up to one according to the detection energy actually delivered, which falls short of the requirement under residual reuse interference or power saturation. $\mathrm{SQH}_i(t)$ is therefore a graded record of delivered sensing performance, bounded in $[0, 1]$. Given $\mathrm{SQH}_i(0) = 0$, a low $\mathrm{SQH}_i$ identifies a vehicle that has been persistently underserved, which raises its priority through~\eqref{eq:SP}.
Integrating historical sensing reliability, safety relevance, and data freshness, we define the VoI for $V_i$ as
\begin{equation}
\mathrm{VoI}_i(t) = \frac{\mathrm{SQH}_i(t) g(\rho_i(t))}{\mathrm{AoI}_i(t)}
\label{eq:VoI}
\end{equation}
\par
Each $\mathrm{VoI}_i(t)$ is bounded in $[0,1]$, with the upper bound approached only when a vehicle is consistently well-served ($\mathrm{SQH}_i\to1$), highly safety-relevant ($g(\rho_i(t))\to1$), and fresh ($\mathrm{AoI}_i(t)=1$). Network-wide performance is evaluated using an $\alpha$-fair aggregate of the per-vehicle VoI
\begin{equation}
\mathcal{U}(t) = \sum_{i \in \mathcal{V}_{\mathrm{sense}}} \big(\eta + \mathrm{VoI}_i(t)\big)^{1-\alpha},
\label{voi_up}
\end{equation}
With $\alpha = 0.5$ and $\eta = 10^{-3}$, the parameter $\alpha$ interpolates between the plain sum ($\alpha = 0$), which is maximized by concentrating service on the highest-relevance vehicles, and, as $\alpha \rightarrow 1$, the proportional-fair (logarithmic) aggregate. Since the sensing budget~\eqref{C1} bounds the total sensing service available per round, the plain sum cannot reward the long-term fairness that SPARC-V is designed to provide, whereas a concave aggregate penalizes schedules that leave vehicles persistently unserved.
\section{Problem Formulation}
The ideal objective is to jointly maximize network-wide VoI as
\begin{equation}
 \max_{\{M,\,\mathbf{P^r}, \mathbf{P^c}, \mathbf{s^r}, \mathbf{s^c}, \mathbf{z} \}} \mathcal{U}(t) 
\label{eq:VoI_max}
\end{equation}
where $M$ is the MCS index vector, $P^r$ 
and $P^c$ are the radar and communication power vectors, $s^r$ and $s^c$ are the sensing and communication mini-slot allocation vectors, and $z \in \{0,1\}^{N\times N}$ is the binary reuse indicator matrix. This constitutes a mixed-integer nonlinear programming (MINLP) problem, which is generally NP-hard due to the interdependence of binary reuse indicators, integer slot assignments, and the non-convex VoI objective. Therefore, we decompose problem~\eqref{eq:VoI_max} into two sequential sub-problems reflecting the safety-first resource hierarchy of ISAC-assisted V2X networks.
\par
To prioritize sensing, we define the SP metric for $V_i$ as
\begin{equation}
\mathrm{SP}_i(t) = \frac{g(\rho_i(t))}{\mathrm{SQH}_i(t) + \varepsilon_s}
\label{eq:SP}
\end{equation}
where $\varepsilon_s$ is a small constant that prevents division by zero when $\mathrm{SQH}_i(t) = 0$ at initialization. A high $\mathrm{SP}_i$ indicates that $V_i$ is safety-critical, historically underserved, or both, so a vehicle with moderate current risk but a poor sensing record can still regain priority. We introduce a binary sensing grant indicator $x_i(t) \in \{0,1\}$, where $x_i(t) = 1$ if $V_i$ has been allocated sensing resources in round \textit{t} ($s^r_i(t) > 0$), and $x_i(t) = 0$ otherwise.
\begin{subequations}
\label{SP}
\begin{align}
\textbf{($\mathcal{P}_1$)}:\quad
& \max_{\{\mathbf{x}, \mathbf{P^r}, \mathbf{s^r}, \mathbf{z} \}} \sum_{i \in \mathcal{V}_{\mathrm{sense}}} x_i \mathrm{SP}_i(t) , \label{SP_OF} \\
\text{s.t.}\quad
&\; \sum_{i \in \mathcal{V}_{\mathrm{sense}}} \hspace {-3mm} s_i^r \le S_{\mathrm{total}}, \ s_i^r \in \mathbb{Z}_{\ge 0}, \ s_i^r \leq S_{CPI}^{\max}, \label{C1} \\
&\; x_i = 1 \iff s_i^r \geq 1,  \label{C3} \\
&\; P^r_i \leq P^r_{\max}  \ ,  \label{C6} \\
&\; z_{i,j} \leq \delta_i \ \delta_j \  \ i \neq j, \ z_{i,j} \in \{0,1\}, \label{C7} \\
&\; z_{i,j}\,\|\mathbf{q}_i-\mathbf{q}_j\|_2\ge z_{i,j}\,d^{\min}_{\mathrm{sep}}, \label{C8}  \\
&\; \gamma^r_i N_i^{sym} \ge \gamma^r_{min} \ \forall \ i \in \mathcal{V}_{\mathrm{sense}}: x_i=1. \label{C10} 
\end{align}
\end{subequations}
\par
where $S^{max}_{CPI}$ is the maximum number of mini-slots in a radar coherent processing interval (CPI), and~\eqref{C10} is the sensing-adequacy target; any shortfall under reuse interference or power saturation is recorded through~\eqref{eq:SQH}. The problem $\mathcal{P}_1$ \eqref{SP} is a weighted selection problem under a knapsack-type budget constraint \eqref{C1}, jointly coupled with the spatial reuse constraints \eqref{C7}–\eqref{C8} and the sensing-adequacy constraint \eqref{C10}.
The residual pool $S_{\mathrm{avail}}$ is what remains of $S_{\mathrm{total}}$ after solving $\mathcal{P}_1$; both~\eqref{C1} and $S_{\mathrm{avail}}$ are evaluated per 1~ms block, where each block hosts one sensing burst, shared by any reusing vehicles, and releases its remaining mini-slots to communication.
\par
Given the residual resource pool $S_{\mathrm{avail}}$, the communication sub-problem $\mathcal{P}_2$ minimizes the network-wide AoI by allocating the mini-slots among communication-feasible vehicles $\mathcal{V}_{\mathrm{comm}}$ in order of information staleness.
\begin{subequations}
\label{VoI_OF}
\begin{align}
\textbf{($\mathcal{P}_2$)}:\quad
& \max_{\{\mathbf{y}, \mathbf{M}, \mathbf{P^c}, \mathbf{s^c} \}} \sum_{i \in \mathcal{V}_{\mathrm{comm}}} y_i \mathrm{AoI}_i(t) , \label{AoI_OF} \\[2mm]
\text{s.t.}\quad
&\; \sum_{i \in \mathcal{V}_{\mathrm{comm}}} s_i^c \le S_{\mathrm{avail}}, \ s_i^c \in \mathbb{Z}_{\ge 0} , \label{C21} \\
&\; y_i = 1 \hspace{-2mm} \iff \hspace{-2mm}  (s_i^c \geq 1) \wedge (\gamma_i^c \geq \gamma_{target}^c(M_i)), \label{C24} \\
&\; M_i \in \mathcal{M}_{\mathrm{3GPP}}  \ , \label{C25}\\
&\; P^c_i \leq P^c_{\max}  \ , \label{C26} 
\end{align}
\end{subequations}
where $\mathcal{M}_{\mathrm{3GPP}}$ is the set of 3GPP NR MCS indices and $y_i$ indicates successful decoding. Since decoded vehicles reset to $\text{AoI}_i(t+1)=1$ and all others increment, $\sum_i \text{AoI}_i(t+1)=K-\sum_i y_i\,\text{AoI}_i(t)$ with $K=\sum_i\big(\text{AoI}_i(t)+1\big)$ independent of $y$, so minimizing post-allocation AoI is equivalent to maximizing~\eqref{AoI_OF}.
\section{Proposed Solution}
\label{sec:proposed}
Since exact solvers for the NP-hard \(\mathcal{P}_1\)~\eqref{SP} and \(\mathcal{P}_2\)~\eqref{VoI_OF} are too costly for a 25~ms scheduling period, SPARC-V solves them sequentially with a greedy heuristic, using \(\mathrm{SP}_i(t)\) and the AoI ordering to obtain high-quality allocations in a single forward pass.
\par
\textbf{Remark 1:} When the per-vehicle mini-slot demand $s_i^r$ is uniform, which holds whenever the radar power constraint \eqref{C6} is not active, the budget constraint \eqref{C1} reduces to a cardinality constraint and greedy selection in descending $SP_i$ order is an exact maximizer of \eqref{SP_OF} over orthogonal assignments. The reuse constraints \eqref{C7}--\eqref{C8} can only admit further vehicles beyond that set, so the greedy allocation attains at least the optimum of the orthogonal-only restriction. Quantifying the residual gap when \eqref{C6} is active, and hence $s_i^r$ varies, is left to future work.
\subsection{Safety-Tiered Sensing Allocation for $\mathcal{P}_1$}
\label{subsec:stage1}
SPARC-V allocates the sensing resource blocks from the available pool $S_{\mathrm{total}}$ to the eligible sensing set $\mathcal{V}_{\mathrm{sense}}\subseteq\mathcal{V}$, subject to the safety-tier and reuse constraints \eqref{C1}--\eqref{C10}. At the start of each round $t$, the RSU computes the risk metric $\rho_i(t)$ using~\eqref{eq:risk} and sets the adaptive risk threshold to the median of all valid risk values, $\rho_{\mathrm{th}}(t)=\operatorname{median}\bigl\{\rho_i(t):i\in\mathcal{V}_{\mathrm{sense}}\bigr\}$.
\par
\textbf{Remark 2:} The threshold $\rho_{th}(t)$ is relative, ensuring that the safety-relevance function $g(\rho_i)$ remains within a dynamically active range, regardless of traffic density and speed. Since $\rho_{th}(t)$ is recomputed from the empirical distribution of risk values at every round rather than fixed a priori, it introduces no tuning parameter: the critical and cautionary tiers adapt automatically to changes in density and relative speed. The constants $\varepsilon_v$ in \eqref{eq:risk} and $\varepsilon_s$ in \eqref{eq:SP} are numerical safeguards against division by zero, set more than an order of magnitude below the typical magnitudes of $v_{rel,i}$ and $SQH_i$ respectively, and therefore affect the orderings only among vehicles whose $v_{rel,i}$ or $SQH_i$ is near zero.

\par
Next, we calculate the SP of \( V_i \) at round \( t \) using \eqref{eq:SP}. The vehicles are then sorted in descending order of \( \mathrm{SP}_i(t) \) to create a scheduling sequence, and resources are assigned sequentially according to this order, following the safety-tier constraints in equations \eqref{C7}--\eqref{C8}. Vehicles without an available resource block are deferred to the next round, so that their SQH falls and their sensing priority \eqref{eq:SP} rises in future rounds.
\par
The minimum radar transmit power \( P_i^{r*} \) is computed for each vehicle. For orthogonal vehicles (\( z_{ij} = 0 \)), there is no interference; however, for reused vehicles (\( z_{ij} = 1 \)), interference occurs.
\begin{equation}
    P_{i,1}^{r*} = \frac{\gamma_{\min,1}^r (\sigma_{\mathrm{VU}}^2 + \sum_{j : z_{ij}=1} P_j^r  \rvert h^{v2v}_{j \rightarrow i } \rvert^{2} )}{|h_i^r|^2},
    \label{eq:P_reuse}
\end{equation}
\par
If $P_i^{r*}$ exceeds the maximum transmit power $P^r_{\max}$, a mini-slot extension mechanism (additional OFDM symbols) is
invoked rather than simply clamping power. The 
SINR target per-symbol is then relaxed to $\gamma_{\min,1}^r = \gamma_{\min}^r /
N_{\mathrm{sym}}^{\mathrm{req}}$ and the final transmit power is $    P_i^{r*} = \min\!\left(
    P_{i,1}^{r*},\;
    P^r_{\max}\right).$
\par
The number of mini-slots consumed by vehicle $i$ is $s_i^r = \lceil
N_{\mathrm{sym}}^{\mathrm{req}} / N_{\mathrm{sym}} \rceil$, following the integration-gain power control of \cite{11594737}. For each block $s$, the ledger records the largest demand $r(s)$ among the vehicles assigned to it, leaving $S_{\mathrm{avail}}(s)=\max(0,N_{mini}-r(s))$ mini-slots for communication.

\subsection{AoI-Driven Communication Allocation for $\mathcal{P}_2$}
\label{subsec:stage2}

$S_{\mathrm{avail}}$ is allocated to uplink CAM
transmissions by a greedy AoI scheduling policy. For each vehicle, communication assignments are orthogonal and the highest feasible MCS index $M_i^*$ is selected as $M_i^*=\max\{M\in\mathcal{M}_{\mathrm{3GPP}}:\gamma_i^c(t)\geq\gamma^c_{target}(M)\}$. If no MCS is feasible, vehicle $i$ is declared in outage for the current scheduling period. Otherwise, the number of physical resource
blocks (PRBs) required to deliver the CAM payload of $L_{CAM}$ bits is
\begin{equation}
n_i^{\mathrm{PRB}}=\left\lceil\frac{L_{CAM}}{12\,N^c_{sym}\,Q_{M_i^*}\,R_{M_i^*}}\right\rceil,
    \label{eq:nPRB}
\end{equation}
where $N^c_{sym}$ is the number of data symbols per mini-slot, and $Q_{M_i^*}$ and $R_{M_i^*}$ are the modulation order and code rate at MCS index $M_i^*$; the CAM then occupies $s^c_i=\lceil n^{PRB}_i/N_{PRB}\rceil$ mini-slots, with $N_{PRB}$ PRBs per mini-slot. 
\par
Vehicles are sorted in descending order of their current AoI to form the communication scheduling order. The scheduler then allocates
mini-slots from $S_{\mathrm{avail}}$ greedily in this staleness
order. A vehicle is successfully decoded if it receives sufficient mini-slots; otherwise, it is declared in outage. The AoI is
then updated as \eqref{eq:AoI2}.
At the end of each round, $\mathrm{SQH}$ is updated via~\eqref{eq:SQH} and the network aggregate $\mathcal{U}(t)$ is evaluated.
\subsection{Computational Complexity}
\label{subsec:complexity}
Per round, the risk metric costs \( O(|\mathcal{V}_{\mathrm{sense}}|) \), sorting by \eqref{eq:SP} costs \( O(|\mathcal{V}_{\mathrm{sense}}| \log |\mathcal{V}_{\mathrm{sense}}|) \), and greedy sensing allocation costs \( O(|\mathcal{V}_{\mathrm{sense}}| S_{\mathrm{total}}) \). Power allocation \eqref{eq:P_reuse} is \( O(|\mathcal{V}_{\mathrm{sense}}|^2) \) in the worst case owing to interference summation, reducing to \( O(|\mathcal{V}_{\mathrm{sense}}|) \) on average since \eqref{C8} limits the number of reuse pairs. The communication stage requires one sort, $O(N \log N)$, and a greedy pass, $O(N S_{\mathrm{total}})$. With \( S_{\mathrm{total}} \) fixed, SPARC-V has worst-case complexity $O(N^2)$ and average-case complexity of $O(N \log N)$. At $N = 100$, the full per-round MATLAB workload (geometry, channel realization, all five schedulers and metric evaluation) executes in 1.21 ms on an Apple M4 laptop, an upper bound on the SPARC-V allocation alone and well within the 25 ms scheduling period.

\begin{figure}[t] 
\centering
\includegraphics[width=0.88\columnwidth]{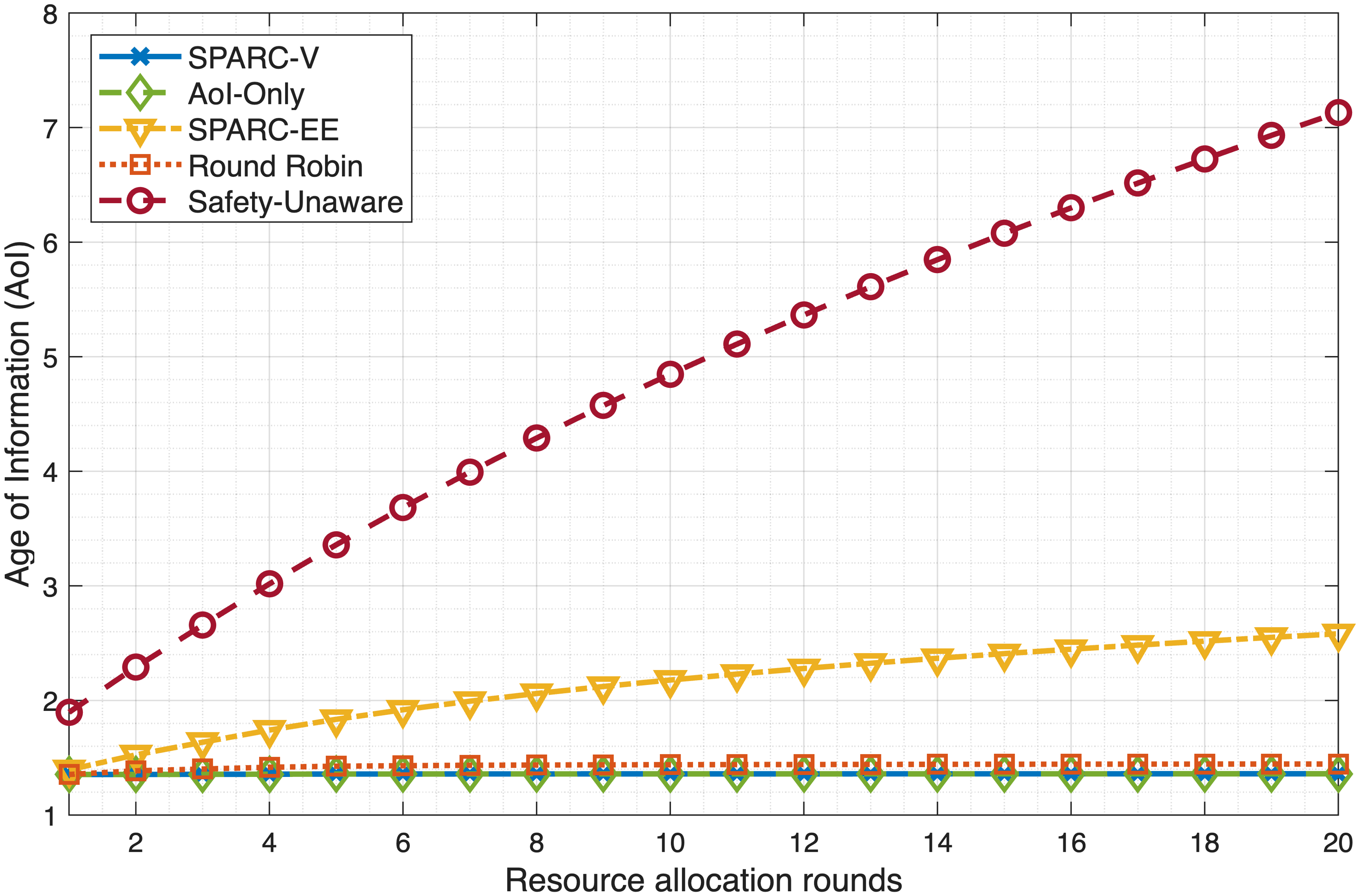}
\caption{Convergence of cumulative mean AoI (N = 100 ISAC vehicles, 500-byte CAM). One scheduling round corresponds to $T_{total}$ = 25 ms.}
\label{AoI}
\vspace{-2.5mm}
\end{figure}
\section{Simulation Results}
We compare SPARC-V against four baselines. The AoI-Only ablation uses the max-age-first communication policy of \cite{AoI}, identically to SPARC-V, but removes SQH from \eqref{eq:SP} so that $\mathrm{SP}_i(t)=g(\rho_i(t))$; it is therefore AoI-neutral and isolates the contribution of the history term. SPARC-EE is our earlier SPARC energy-efficiency scheduler \cite{11594737}, which enforces strict sensing orthogonality with no reuse and base-packs communication users in ascending order of channel quality. Round Robin serves both stages cyclically, advancing one full resource window per round. Safety-Unaware removes safety tiering and assigns sensing and communication slots at random without the separation constraint, so interference is uncontrolled in both stages.
\par
Vehicles are deployed on a road of length $L = 1000$ m in two lanes $5$ m apart, with the RSU at $q_{RSU} = [500, 10]$ m, and follow a first-order autoregressive mobility model with mean speed $v_r = 14$~m/s and standard deviation $\sigma_v = 2$~m/s. The system operates at $f_c = 5.9$~GHz over $B = 70$~MHz, with $d_{max}^r = 100$ m, $d_{sep}^{min} = 250$ m, $\sigma_{\chi} = 3$ dB and $P_{NLoS} = 0.2$. Radar parameters follow the Neyman-Pearson framework with \( P_d = 0.95 \), \( P_{\mathrm{fa}} = 10^{-4} \) and $\sigma_r=15$~m$^2$. Maximum transmit powers are $P^r_{\max}=1$~W and $P^c_{\max}=0.2$~W, noise  figures are $F_{VU}=7$~dB and $F_{RSU}=4$~dB, and MCS selection follows the 3GPP NR table \cite{3GPPMCS}. By default, each sensing burst uses 35 symbols (5 mini-slots); $N^c_{sym}=4$, $N_{PRB}=93$, vehicles are placed at least 20~m apart, and $G=5$~dBi, $G_c=15$~dB.
\begin{figure}[t] 
\centering
\includegraphics[width=0.88\columnwidth]{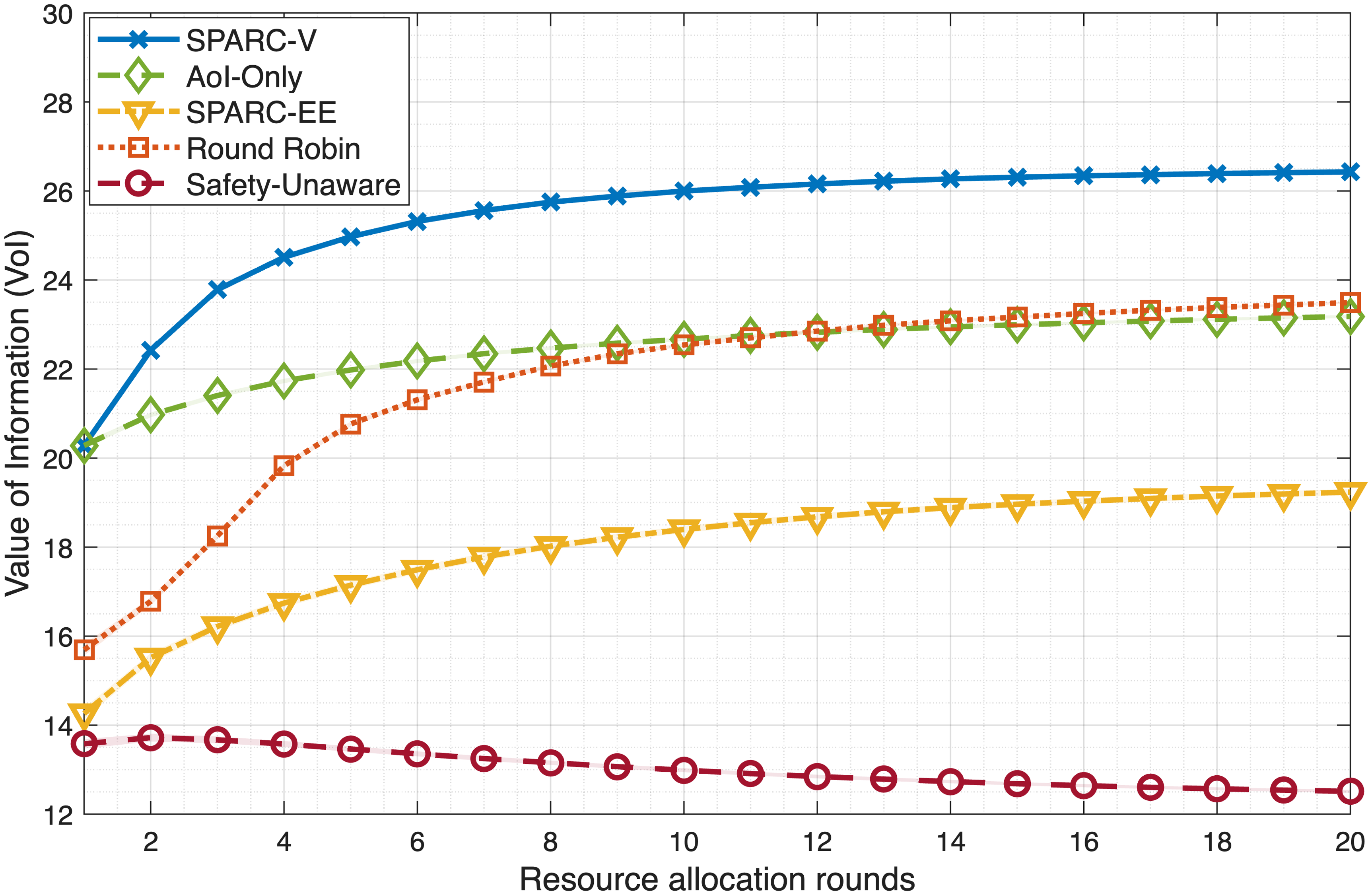}
\caption{Convergence of cumulative network VoI (N = 100 ISAC vehicles, 500-byte CAM). Shaded 95\% confidence intervals over 500 Monte Carlo trials are narrower than the line width.}
\label{VoI}
\vspace{-5mm}
\end{figure}
\par
\begin{figure}[t] 
\centering
\includegraphics[width=0.88\columnwidth]{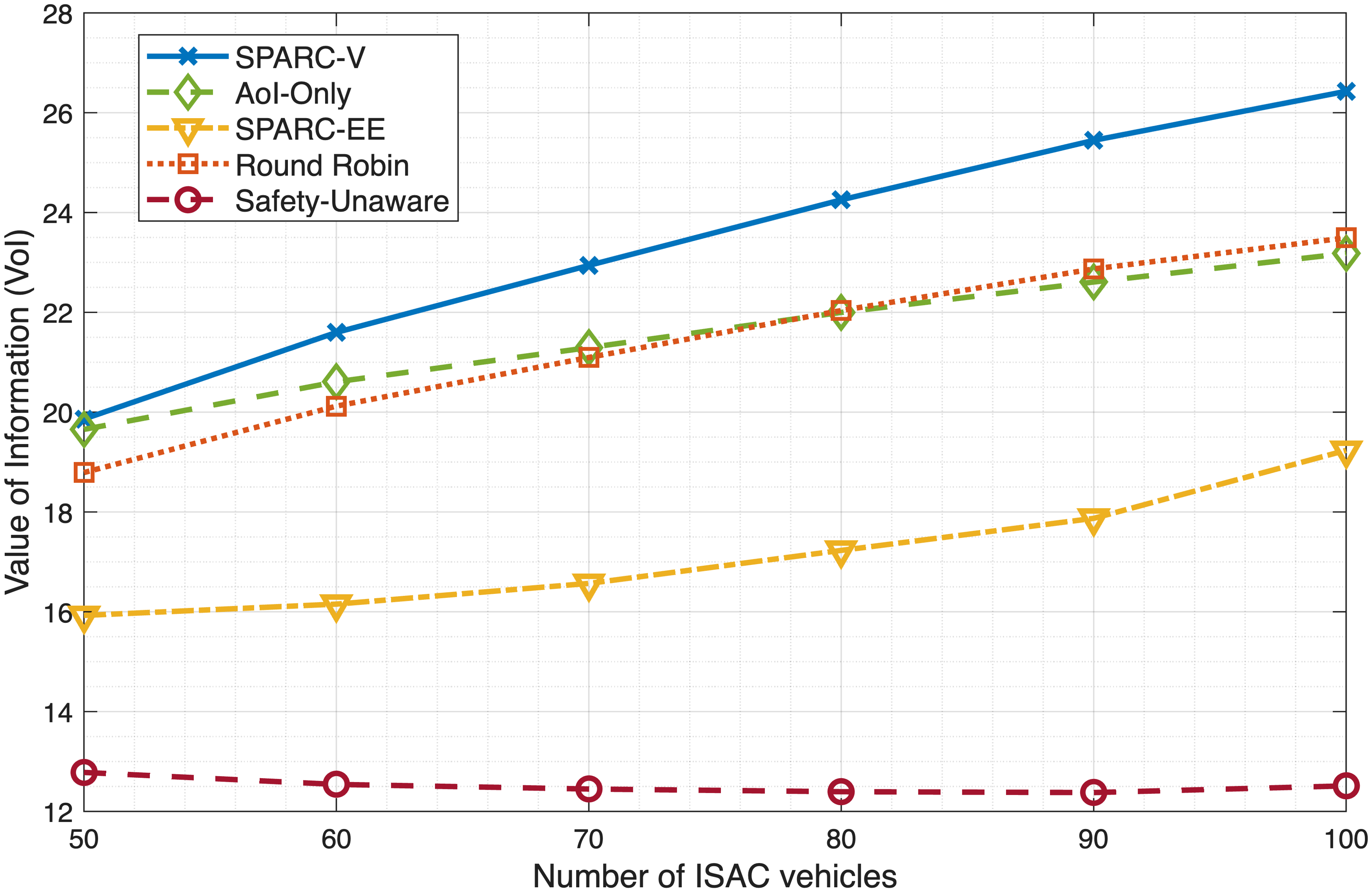}
\caption{Scalability of cumulative network VoI with the number of ISAC vehicles (500-byte CAM, fixed resource pool).}
\label{N_voi}
\vspace{-5mm}
\end{figure}
\begin{figure}[t] 
\centering
\includegraphics[width=0.88\columnwidth]{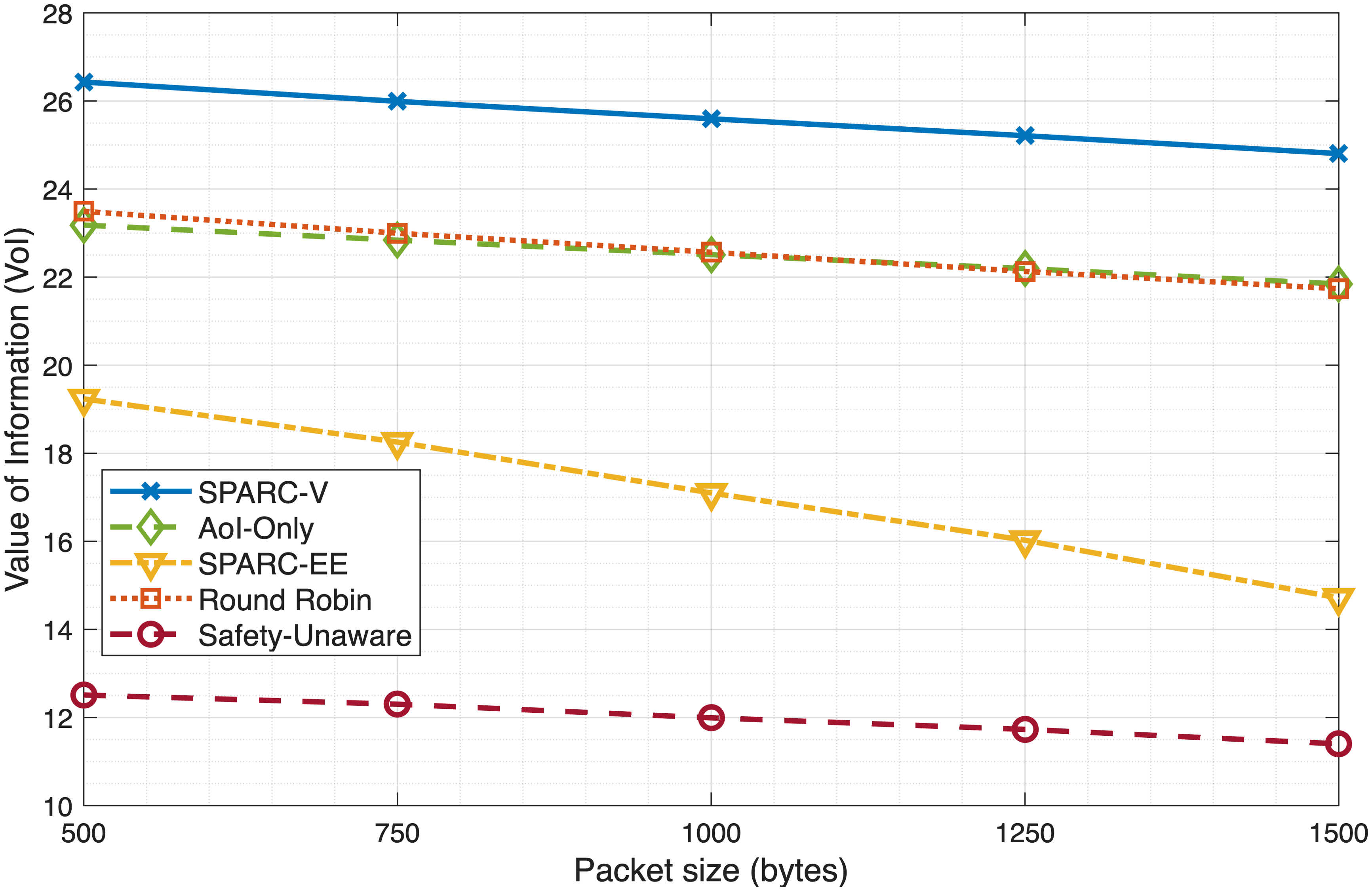}
\caption{Sensitivity of cumulative network VoI to CAM payload size (N = 100 ISAC vehicles). The 500-byte nominal payload used in Figs. 2 to 4 is the leftmost point.}
\label{camsize_voi}
\vspace{-2.5mm}
\end{figure}

Figs.~\ref{AoI} and~\ref{VoI} show the cumulative mean AoI and cumulative network VoI over the scheduling rounds for 100 ISAC vehicles and a 500-byte CAM, averaged over 500 Monte Carlo trials. Across every scheme and configuration in Figs.~\ref{AoI} to~\ref{camsize_voi}, the 95\% confidence half-width does not exceed 0.10 on the cumulative network VoI and 0.04 on the cumulative mean AoI, so the orderings reported below are statistically unambiguous. The nominal payload is representative of ETSI CAMs, which typically range up to about 1~kB; Fig.~\ref{camsize_voi} extends to 1500 bytes as an upper stress point. Mean AoI exceeds its initial value of 1 from the first round, since the residual pool cannot serve every vehicle in a single round.
\par
In Fig.~\ref{AoI}, the schemes separate by how they manage the communication stage. SPARC-V and the AoI-Only ablation coincide exactly at 1.36 rounds (34~ms), since they share the same communication scheduler; the effect of the SQH term therefore appears in Fig.~\ref{VoI}. Round Robin reaches 1.45 rounds, as its cyclic ordering serves every vehicle but ignores staleness, and SPARC-EE rises to 2.58 rounds because weakest-first ordering repeatedly defers vehicles that are already stale. Safety-Unaware degrades most severely, reaching 7.13 rounds (178~ms), as co-channel interference at the RSU drives vehicles into outage regardless of their staleness.
\par
Fig. \ref{VoI} shows that SPARC-V attains the highest cumulative network VoI at 26.43, exceeding the closest baseline, Round Robin, by 12.5\%, a margin far exceeding the 95\% confidence half-widths of 0.025 for both schemes. At the first round SPARC-V and the AoI-Only ablation coincide: with $\mathrm{SQH}_i(0)=0$, the priority in \eqref{eq:SP} reduces to $g(\rho_i)/\varepsilon_s$, which orders vehicles identically to the ablation. The schemes are otherwise identical, so the 14\% separation by round 20 is attributable entirely to the accumulated sensing history. At the final round, SPARC-V's per-vehicle SQH has mean 0.261, standard deviation 0.104 and range 0.020 to 0.600, so the history term separates vehicles by accumulated service rather than saturating at a common value; its Jain fairness index is 0.864 against 0.404 for the ablation, which leaves 33.1\% of eligible vehicles unsensed over the 20 rounds while SPARC-V leaves none. Round Robin and AoI-Only converge near 23 and SPARC-EE to 19.24, while Safety-Unaware is the only scheme whose VoI declines, from 13.60 to 12.51, as unprotected reuse degrades sensing quality cumulatively. Under the plain sum ($\alpha = 0$), by contrast, the ablation leads at $12.12$ against $10.07$; a binding budget \eqref{C1} bounds the total sensing service available per round, so the plain sum is maximized by concentrating it on the highest-relevance vehicles, at the cost of the starvation noted above.
\par
Fig.~\ref{N_voi} varies the number of ISAC vehicles from 50 to 100 with a fixed resource pool. SPARC-V rises from 19.95 to 26.43, partly because more vehicles contribute to the aggregate, but its margin over the strongest baseline, which isolates relative efficiency, widens monotonically from 1.3\% at 50 vehicles to 12.5\% at 100. As density grows, a smaller fraction of the fleet can be sensed per round, and the history term is what prevents persistent deprivation. The strongest baseline also changes with density. AoI-Only leads below 80 vehicles, where safety weighting matters most, and Round Robin above, where uniform coverage matters more; SPARC-V exceeds both throughout, confirming that neither safety weighting nor fairness alone suffices. Safety-Unaware stays near 12.5 at every density, limited by uncoordinated reuse.
\par
Fig.~\ref{camsize_voi} varies the CAM payload from 500 to 1500 bytes. All schemes decline as payloads grow, since each transmission consumes more of the residual mini-slot pool. SPARC-V leads at every payload by a margin far exceeding the confidence intervals, falling 6.1\% from 26.43 to 24.83 while retaining an approximately constant absolute margin, so its advantage does not depend on the choice of CAM size. The baselines decline by roughly 6 to 9\%, except SPARC-EE, which falls 23.3\% from 19.24 to 14.75; serving the weakest links first lets them consume a growing share of the pool as payloads increase. AoI-Only and Round Robin cross near 1000 bytes, where they are statistically indistinguishable.
\section{Conclusion}
This paper presented SPARC-V, a low-complexity two-stage ISAC-V2X scheduler whose history-aware VoI metric raises the priority of vehicles that are safety-critical but historically underserved. SPARC-V exceeds the strongest baseline by 12.5\% in network VoI at 100 vehicles, a margin that widens with density and holds across CAM payloads, and leaves no eligible vehicle unsensed over the evaluation horizon, against 33.1\% for the history-free ablation. Future work will quantify the optimality gap of the greedy allocation when the radar power constraint is active.
\bibliographystyle{IEEEtran}
\bibliography{Ref_Paper_1}

\begin{thebibliography}{10}
\providecommand{\url}[1]{#1}
\csname url@samestyle\endcsname
\providecommand{\newblock}{\relax}
\providecommand{\bibinfo}[2]{#2}
\providecommand{\BIBentrySTDinterwordspacing}{\spaceskip=0pt\relax}
\providecommand{\BIBentryALTinterwordstretchfactor}{4}
\providecommand{\BIBentryALTinterwordspacing}{\spaceskip=\fontdimen2\font plus
\BIBentryALTinterwordstretchfactor\fontdimen3\font minus \fontdimen4\font\relax}
\providecommand{\BIBforeignlanguage}[2]{{%
\expandafter\ifx\csname l@#1\endcsname\relax
\typeout{** WARNING: IEEEtran.bst: No hyphenation pattern has been}%
\typeout{** loaded for the language `#1'. Using the pattern for}%
\typeout{** the default language instead.}%
\else
\language=\csname l@#1\endcsname
\fi
#2}}
\providecommand{\BIBdecl}{\relax}
\BIBdecl

\bibitem{zhang2026integrated}
D.~Zhang, Y.~Cui, X.~Cao, N.~Su, Y.~Gong, F.~Liu, W.~Yuan, X.~Jing, J.~A. Zhang, J.~Xu \emph{et~al.}, ``Integrated sensing and communications over the years: An evolution perspective,'' \emph{IEEE Commun. Surveys Tuts.}, 2026.

\bibitem{11219062}
B.~Liu, H.~Shi, D.~Jia, E.~Wang, W.~Han, K.~Zhong, L.~Wu, S.~Chen, C.~Qiao, and J.~Wang, ``Collaborative sensing and communication for intelligent connected vehicles: A comprehensive survey,'' \emph{IEEE Commun. Surveys Tuts.}, vol.~28, pp. 3125--3164, 2026.

\bibitem{11396674}
J.~Qi, C.~Liu, C.~Xu, R.~Wang, J.~Dong, and Y.~Yu, ``Efficient information updates in compute-first networking via reinforcement learning with joint {AoI} and {VoI},'' \emph{IEEE Internet Things J.}, vol.~13, no.~9, pp. 19\,699--19\,711, 2026.

\bibitem{AoI}
I.~Kadota, A.~Sinha, E.~Uysal-Biyikoglu, R.~Singh, and E.~Modiano, ``Scheduling policies for minimizing age of information in broadcast wireless networks,'' \emph{IEEE/ACM Trans. Netw.}, vol.~26, no.~6, pp. 2637--2650, 2018.

\bibitem{wang2024TVT}
W.~Wang \emph{et~al.}, ``Value matters: A novel value of information-based resource scheduling method for {CAV}s,'' \emph{IEEE Trans. Veh. Technol.}, vol.~73, no.~6, pp. 8720--8735, 2024.

\bibitem{rolich2026latency}
A.~Rolich, M.~Yildiz, I.~Turcanu, A.~Vinel, and A.~Baiocchi, ``From latency to value of information: A review of timeliness metrics for safe transportation systems in the {6G} era,'' \emph{IEEE Veh. Technol. Mag.}, 2026.

\bibitem{liu2020radar}
F.~Liu, W.~Yuan, C.~Masouros, and J.~Yuan, ``Radar-assisted predictive beamforming for vehicular links: Communication served by sensing,'' \emph{IEEE Trans. Wireless Commun.}, vol.~19, no.~11, pp. 7704--7719, 2020.

\bibitem{11594737}
M.~Sarfraz, N.~Imani, S.~Lavdas, and C.~B. Papadias, ``Safety-centric resource scheduling for energy efficiency optimization in {ISAC}-assisted {V2X},'' in \emph{Proc. Int. Conf. Telecommun. (ICT)}, 2026, pp. 43--48.

\bibitem{abedi2026precise}
M.~R. Abedi, Z.~Rashidi, N.~Mokari, H.~Saeedi, and N.~Zorba, ``Precise {HDV} positioning through safety-aware {ISAC} in a value-of-information-driven {6G} {V2X} system,'' \emph{IEEE Trans. Wireless Commun.}, vol.~25, pp. 14\,875--14\,891, 2026.

\bibitem{3GPPNLoS}
``Study on evaluation methodology of new vehicle-to-everything ({V2X}) use cases for {LTE} and {NR},'' 3GPP, TR 37.885, V15.3.0, Jun. 2019.

\bibitem{de2008design}
A.~De~Maio, S.~De~Nicola, Y.~Huang, Z.-Q. Luo, and S.~Zhang, ``Design of phase codes for radar performance optimization with a similarity constraint,'' \emph{IEEE Trans. Signal Process.}, vol.~57, no.~2, pp. 610--621, 2009.

\bibitem{3GPPMCS}
``{NR}; physical layer procedures for data,'' 3GPP, TS 38.214, V16.2.0, Jun. 2020.

\bibitem{3GPPalpha}
``Evolved universal terrestrial radio access ({E-UTRA}); radio frequency ({RF}) system scenarios,'' 3GPP, TR 36.942, V13.0.0, Jan. 2016.

\end{thebibliography}
\end{document}